\documentclass{optica-article}

\journal{opticajournal} 

\articletype{Research Article}

\usepackage{soul}
\usepackage{multicol}
\usepackage{appendix}
\usepackage{orcidlink}
\usepackage{lineno}
\usepackage{fancyhdr}

\begin{document}

\title{Large-field-of-view lensless imaging with miniaturized sensors}

\author{Yu Ren\authormark{1}\orcidlink{0000-0002-8582-680X}, Xiaoling Zhang\authormark{2}\orcidlink{0000-0003-2343-3055}, Xu Zhan\authormark{2}\orcidlink{0000-0003-2816-9791}, Xiangdong Ma\authormark{2}\orcidlink{0000-0002-2834-2163}, Yunqi Wang\authormark{3}\orcidlink{0000-0002-1584-5132}, Edmund Y. Lam\authormark{4}\orcidlink{0000-0001-6268-950X}, and Tianjiao Zeng\authormark{1,*}\orcidlink{0000-0002-6780-6100}}

\address{\authormark{1}School of Aeronautics and Astronautics, University of Electronic Science and Technology of China, Chengdu 611731, China \\
\authormark{2}School of Information and Communication Engineering, University of Electronic Science and Technology of China, Chengdu 611731, China\\
\authormark{3}Psychiatry Neuroimaging Laboratory, Brigham and Women’s Hospital, Harvard Medical School, Boston, MA 02215, United States of America\\
\authormark{4} Department of Electrical and Electronic Engineering, The University of Hong Kong, Pokfulam, HongKong, China
}

\email{\authormark{*}corresponding author:~tzeng@uestc.edu.cn} 

\begin{abstract*}
Lensless cameras replace bulky optics with thin modulation masks, enabling compact imaging systems. However, existing methods rely on an idealized model that assumes a globally shift-invariant point spread function (PSF) and sufficiently large sensors. In reality, the PSF varies spatially across the field of view (FOV), and finite sensor boundaries truncate modulated light—effects that intensify as sensors shrink, degrading peripheral reconstruction quality and limiting the effective FOV. We address these limitations through a local-to-global hierarchical framework grounded in a locally shift-invariant convolution model that explicitly accounts for PSF variation and sensor truncation. Patch-wise learned deconvolution first adaptively estimates local PSFs and reconstructs regions independently. A hierarchical enhancement network then progressively expands its receptive field—from small patches through intermediate blocks to the full image—integrating fine local details with global contextual information. Experiments on public datasets show that our method achieves superior reconstruction quality over a larger effective FOV with significantly reduced sensor sizes. Under extreme miniaturization—sensors reduced to 8\%of the original area—we achieve improvements of 2 dB (PSNR) and 5\% (SSIM), with particularly notable gains in structural fidelity. Code is available at \url{https://github.com/KB504-public/l2g_lensless_imaging}.
\end{abstract*}

\section{Introduction}

Lensless imaging systems substitute the conventional lens—typically the bulkiest component—with a thin modulation mask, achieving compact, lightweight designs~\cite{2017-flatcam_camera, 2018-diffusercam_camera, 2020-phlatcam_camera, 2018-lensless_3D_imaging, 2019-miniature,2019-single_shot, Monakhova:21, Liu:251, Liu:25, Chen:25, Li:25, doi:10.1126/sciadv.adt3909}. Unlike traditional cameras that map each scene point directly to a sensor location, lensless systems encode spatial information into a distributed intensity pattern—the Point Spread Function (PSF)~\cite{2016-lensless_imaging, 2016-computational_renaissance, 2022-review}. The sensor thus captures blurred, unrecognizable patterns rather than sharp images, requiring computational algorithms to invert this encoding and recover interpretable visual content.

Recent advances in deep learning have substantially improved lensless imaging reconstruction~\cite{11251013, 10811979, 2019-leadmm, 2019-pre_flatnet, 2021-mmcn, Kingshott:22, Pan:22, 2022-flatnet, 2023-mwdn, Zhang:23, 2025-modular_recon, 2024-deeplir, 2024-phocolens, 2025-generative}. Current methods adopt three architectural paradigms: two-stage architectures (e.g., FlatNet~\cite{2022-flatnet}) that decompose reconstruction into learned deconvolution and enhancement; deep unfolding architectures (e.g., Le-ADMM~\cite{2019-leadmm} and DeepLIR~\cite{2024-deeplir}) that unroll iterative optimization into trainable layers with learned priors; and multi-scale architectures (e.g., MWDN~\cite{2023-mwdn}) that fuse feature pyramids to integrate semantic context with fine spatial detail. Despite progressively improving reconstruction quality, these frameworks rely on idealized modeling assumptions that constrain performance under practical conditions.

Despite their impressive performance, existing lensless imaging methods rely on idealized assumptions that fail under practical conditions. Most approaches adopt a global convolution framework~\cite{2019-leadmm, 2019-pre_flatnet, 2022-flatnet, 2023-mwdn, 2024-deeplir}, treating the PSF as shift-invariant across the entire FOV and assuming sensors are effectively infinite. In reality, the PSF exhibits slow spatial variation that no single global kernel can capture~\cite{2021-mmcn, 2025-modular_recon, 2024-phocolens}, while finite sensor boundaries truncate substantial portions of the modulated light distribution, causing severe information loss at the periphery.

\textbf{Problems.} Spatial PSF variation and measurement truncation jointly degrade reconstruction quality, particularly compromising structural fidelity in contours, edges, and fine details. Our experiments demonstrate that representative methods across all major architectural paradigms inadequately compensate for these effects, substantially reducing the effective FOV and undermining the large-FOV advantages inherent to lensless imaging. Existing approaches face an unavoidable trade-off: accept diminished usable FOV or employ larger sensors—both contradicting the core miniaturization objectives of lensless systems.

\textbf{Motivations.} To realize the full potential of compact, large field-of-view lensless imaging, this work aims to extend the effective FOV under constrained—or even substantially reduced—sensor dimensions while enhancing reconstruction fidelity, particularly for fine structural details in peripheral regions.

\textbf{Contributions.} To capture the spatial variations of PSF and trunction effects faithfully, we propose a locally shift-invariant convolution model that decomposes the scene into patches, each characterized by its own locally accurate PSF, with truncation explicitly modeled through windowing. Building on this refined measurement model, we introduce a two-stage local-to-global reconstruction framework. The first stage performs patch-wise learned deconvolution: a lightweight network adaptively estimates local PSFs and reconstructs each region independently, respecting local shift-invariance while accounting for truncation. The second stage employs a hierarchical enhancement network that progressively expands its receptive field—from small patches through intermediate blocks to the full image—integrating fine-grained local features with global contextual information to achieve high-fidelity reconstruction with preserved structural detail.

\textbf{Results.} We have validated our approach on two public datasets—DiffuserCam~\cite{2019-leadmm} and PhlatCam~\cite{2022-flatnet}—benchmarking against representative methods across major architectural paradigms~\cite{2022-flatnet, 2024-deeplir, 2023-mwdn}. Our framework consistently outperforms existing methods, achieving superior reconstruction quality across a substantially larger effective FOV even under significantly reduced sensor dimensions. On DiffuserCam at full sensor size, our method yields gains of 0.5–5 dB (PSNR) and 1.5–10\% (SSIM); when sensor area contracts to ~65\% of the original, these advantages grow to 1–5 dB (PSNR) and 3–23\% (SSIM). The PhlatCam dataset exhibits similar trends, with improvements of 1–3 dB (PSNR) and 4–7\% (SSIM) at full sensor size. Most notably, even under extreme miniaturization—where sensor areas shrink to 1/4 or 1/12 of the original—our framework maintains robust gains of 2–4 dB (PSNR) and 5–11\% (SSIM).

\section{Problem Formation}

Most existing methods model lensless imaging as a convolution:
\begin{equation}
\mathbf{Y} = \mathbf{H} \circledast \mathbf{X} + \mathbf{N}
\end{equation}
where $\mathbf{Y} \in \mathbb{R}^{M \times N}$ is the sensor measurement, $\mathbf{X} \in \mathbb{R}^{P \times Q}$ the scene, $\mathbf{H} \in \mathbb{R}^{K \times L}$ the PSF, $\mathbf{N} \in \mathbb{R}^{M \times N}$ sampling noise, and $\circledast$ denotes convolution. This linear shift-invariant (LSI) formulation implicitly assumes two conditions:
\begin{itemize}
\item PSF Spatial Invariance. It presumes that the PSF remains spatially invariant across the entire FOV, such that $\mathbf{H}_{(x,y)} = \mathbf{H}$ for all positions $(x,y)$.
\item Complete Sampling.  It assumes complete sampling, meaning that the support of the convolved measurement $\mathbf{H} \circledast \mathbf{X}$ lies entirely within the sensor's effective sampling area $A_{\text{sensor}}$, i.e., $\text{supp}(\mathbf{H} \circledast \mathbf{X}) \subseteq A_{\text{sensor}}$.
\end{itemize}

\begin{figure}[h]
  \centering
  \includegraphics[width=8cm]{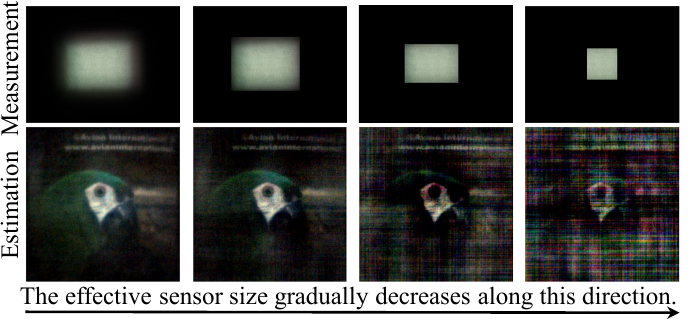}
  \caption{Progressive sensor truncation degrades reconstruction fidelity and confines the effective field of view to the central region. The leftmost panel retains full measurements; the rightmost shows extreme truncation with peripheral information nearly eliminated. All reconstructions use Wiener filtering~\cite{2002-digital}.}
\label{fig:effect_of_sensor_truncation}
\end{figure}

However, these assumptions break down under conditions where lensless imaging offers its greatest advantages: large field-of-view capture with miniaturized sensors. As the FOV expands, the PSF exhibits non-negligible angular dependence—PSFs corresponding to different incident angles vary measurably, as documented in prior studies~\cite{2021-mmcn,2025-modular_recon}. Applying a single, globally invariant PSF across all scene points introduces systematic error that degrades structural fidelity, compromising edge sharpness and fine detail.

More critically, when sensor dimensions shrink to leverage miniaturization benefits, finite sensor boundaries truncate substantial portions of the modulated light distribution. Mathematically, when $M < (P + K - 1)$ and $N < (Q + L - 1)$—or in extreme cases where $M \approx P$ and $N \approx Q$—this truncation causes significant information loss, especially near the image periphery. The issue compounds when pursuing large FOV imaging with compact sensors, where both challenges converge: truncation-induced artifacts severely compromise structural fidelity, causing edges to blur, contours to degrade, and fine details to vanish (Figure~\ref{fig:effect_of_sensor_truncation}), thus reducing effective FOV size.

\section{Proposed Method}

\subsection{Locally convolutional measurement model}

We partition the scene into $B_x \times B_y$ patches ${\{A_b\}}_{b=1}^{B_x \times B_y}$, which are the fundamental processing units, and model the measurement as their superposition. Under local shift-invariance, each patch $A_b$ contributes:

\begin{equation}
\mathbf{Y}_b = \mathcal{W}(\mathbf{H}_b \circledast \mathbf{X}_b, \Omega) + \mathbf{N}_b
\end{equation}
where $\mathbf{H}_b$ denotes the locally accurate shift-invariant PSF for patch $A_b$, $\Omega$ represents the sensor's effective sampling region, and $\mathcal{W}$ is a window function that models sensor truncation:

\begin{equation}
\mathcal{W}(x, \Omega) =
\begin{cases}
1, & x \in \Omega \\
0, & \text{otherwise}
\end{cases}
\end{equation}

The complete measurement is then obtained by superposing the contributions from all patches:

\begin{equation}
    \mathbf{Y} = \sum_{b=1}^{B_x \times B_y} \mathbf{Y}_b = \mathcal{W}\left[ \sum_{b=1}^{B_x \times B_y} (\mathbf{H}_b \circledast \mathbf{X}_b), \Omega \right] + \mathbf{N}
\end{equation}

\subsection{Local-to-global hierarchical reconstruction}

Building on the locally convolutional measurement model, a natural reconstruction strategy is to process each patch independently and combine the results. As derived in the appendix, this can be expressed as:
\begin{equation}
\hat{\mathbf{X}} = \sum_{b=1}^{B_x \times B_y} \left( \mathbf{1}_b \odot \hat{\mathbf{X}}_b \right) = \sum_{b=1}^{B_x \times B_y} \left( \mathbf{1}_b \odot \mathcal{F}^{-1} [w_b \odot \mathcal{F}(\mathbf{Y})] \right)
\label{eq:imaging_model}
\end{equation}
where $\mathbf{1}_b$ is the spatial selection function for patch $b$ (unity within the patch, zero elsewhere), and $w_b$ is the patch-specific deconvolution kernel reflecting local PSF characteristics. However, this approach faces two fundamental challenges: accurately determining the spatially varying kernels $w_b$, and mitigating boundary artifacts that arise when independently processed patches are stitched together, which compromise global structural coherence. We address these limitations through a dual-stage local-to-global reconstruction framework. 

\subsubsection{Patch-wise learning deconvolution}

The first stage performs patch-wise learned deconvolution to extract an initial reconstruction from raw measurements. Leveraging local shift-invariance within small regions, we partition the imaging area into patches, each processed independently via a lightweight neural network that predicts a frequency-domain deconvolution kernel tailored to local PSF characteristics. After filtering the measurement in Fourier space and inverse transformation, a learned normalization coefficient scales the spatial-domain result to complete patch-level deconvolution.

\begin{figure}[h]
  \centering
  \includegraphics[width=12cm]{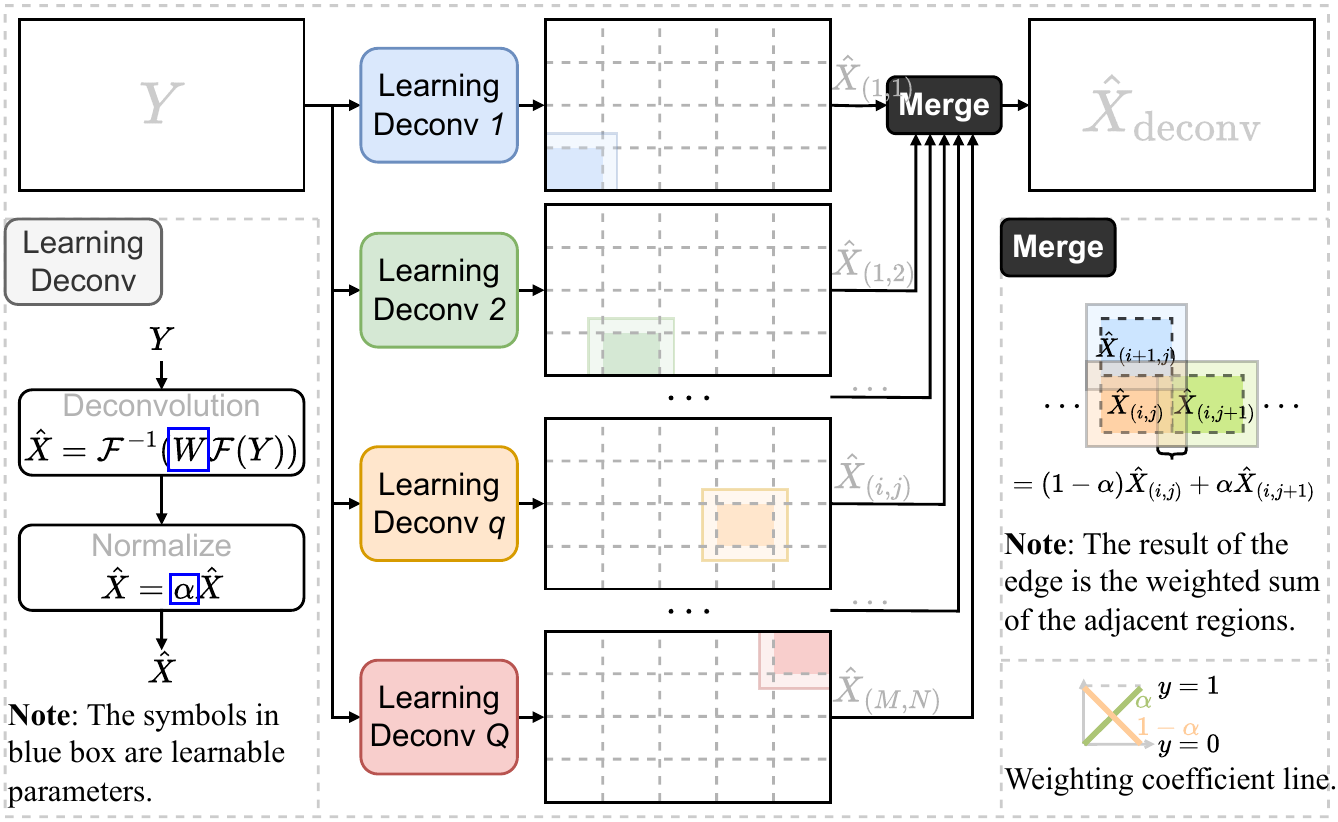}
  \caption{Structure of patch-wise learning deconvolution network. \textit{(To ensure this diagram is clear enough, we used a 4$\times$5 partitioning strategy for demonstration.)}}
\label{fig:deconv_architecture}
\end{figure}

Direct stitching of these independently processed patches would introduce boundary artifacts—intensity discontinuities and structural misalignment—due to differing kernels applied to neighboring regions. To ensure seamless integration, we adopt an overlapping patch strategy: each patch extends slightly beyond its nominal boundaries, creating overlap zones where pixels are fused via linear weighting that smoothly transitions contributions from one patch to the next. Empirical validation confirms that 10–20 pixel overlap widths effectively suppress artifacts without substantial computational overhead. The overall process is depicted in Figure~\ref{fig:deconv_architecture}.

\subsubsection{Local-to-global hierarchical enhancement}

The first stage applies a locally shift-invariant model through a lightweight architecture to transform raw measurements into preliminary reconstructions. While effective at recovering local features, this design has three inherent limitations: patch-wise processing cannot integrate global contextual information across the full FOV; the shallow network lacks capacity to capture high-level semantic features essential for perceptual fidelity; and the framework does not explicitly model degradation factors such as noise. Consequently, first-stage outputs exhibit residual noise, artifacts, and structural inconsistencies, motivating a second refinement stage that operates in the natural image domain.

\begin{figure}[h]
  \centering
  \includegraphics[width=13.3cm]{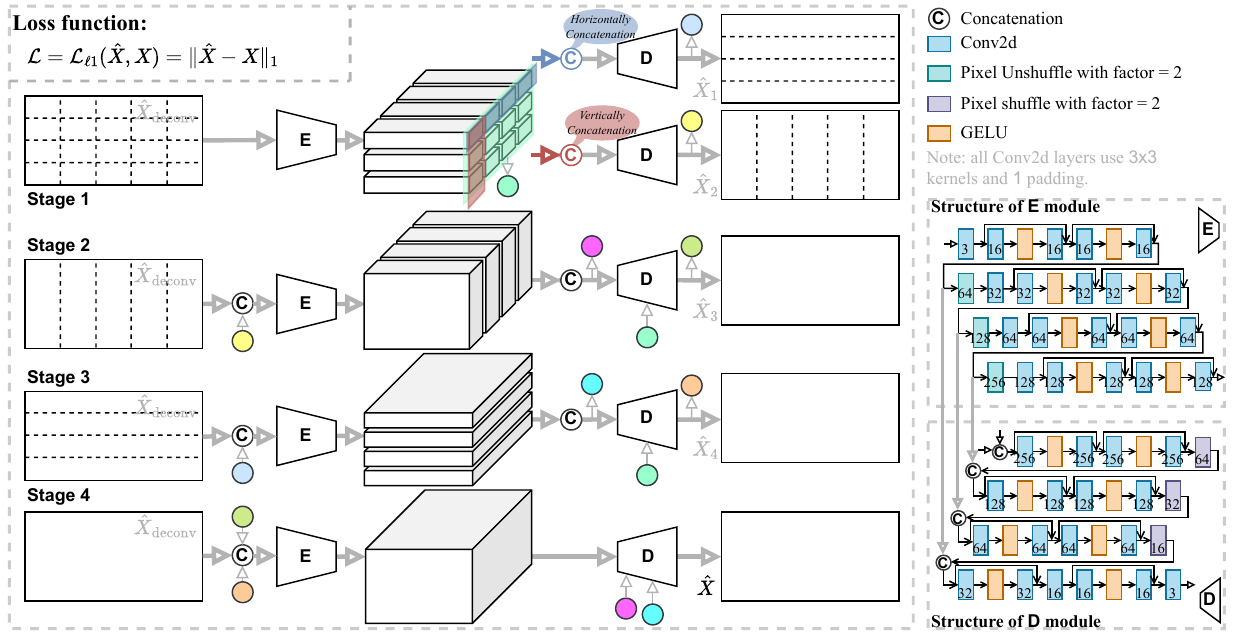}
  \caption{Structure of the network of the enhancement part. \textit{(To ensure this diagram is clear enough, we used a 4$\times$5 partitioning strategy for demonstration.)}}
\label{fig:enhancement_architecture}
\end{figure}

The second stage addresses these deficiencies through hierarchical local-to-global enhancement guided by multi-scale patch features. Rather than simple spatial decoupling, it implements scale-aware semantic processing that integrates local detail with global context via a bottom-up feature propagation pathway, ensuring fine-grained information informs coarse-scale reasoning. The enhancement network implements this strategy through a multi-scale architecture (Figure~\ref{fig:enhancement_architecture}) that partitions input into overlapping patches at progressively coarser scales—individual patches, vertical blocks, horizontal blocks, and the full image—each processed by a dedicated convolutional encoder-decoder. Cross-scale integration occurs through feature concatenation: finer-scale encodings are concatenated with coarser-scale features before decoding, allowing high-level semantic reasoning to leverage detailed local information.

\subsubsection{Loss function}

Both stages of the proposed network are trained using the L1 loss function, which measures the absolute difference between the network's output and the ground truth:
\begin{equation}
\mathcal{L}_{\ell 1} = \left\lVert \hat{\mathbf{X}} - \mathbf{X} \right\rVert_1
\end{equation}
Here, $\hat{\mathbf{X}}$ represents the network-reconstructed image, while $\mathbf{X}$ denotes the corresponding ground-truth reference.

\section{Experiments}

\subsection{Datasets}

We validate our method using two publicly available datasets: DiffuserCam~\cite{2019-leadmm} and PhlatCam~\cite{2022-flatnet}, both constructed by imaging monitor-displayed content through their respective lensless systems to enable controlled comparison between measurements and ground-truth references. To assess whether our approach maintains high-fidelity reconstruction across large effective FOVs under sensor miniaturization, we simulate progressive size reduction through measurement truncation, testing whether our method preserves image quality throughout the full FOV—especially in peripheral regions—without requiring the FOV reduction characteristic of existing approaches.

\begin{figure}[h]
  \centering
  \includegraphics[width=12cm]{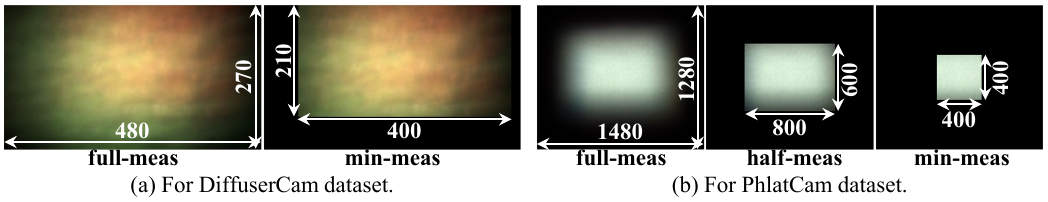}
  \caption{Demonstration of measurement sizes of different experiments.}
\label{fig:meas_size}
\end{figure}

The DiffuserCam dataset contains 25,000 sample pairs (24,000 training, 1,000 evaluation), each pairing a $270 \times 480$ pixel lensless measurement with its corresponding ground-truth image. We conduct two configurations (Figure~\ref{fig:meas_size}a): \textbf{full-meas} retains complete measurement information, while \textbf{min-meas} truncates to the central $210 \times 400$ pixel region—matching the ground-truth footprint and approximating 65\% sensor area reduction. The PhlatCam dataset comprises 10,000 sample pairs spanning 1,000 object categories (990 training, 10 test categories), each pairing a $1518 \times 2012$ pixel measurement with a $384 \times 384$ pixel ground-truth reference. After cropping to $1280 \times 1480$ pixels to remove uninformative boundaries, we evaluate three configurations (Figure~\ref{fig:meas_size}b): \textbf{full-meas} at $1280 \times 1480$ pixels; \textbf{half-meas} at $600 \times 800$ pixels (~25\% of original sensor area); and \textbf{min-meas} at $400 \times 400$ pixels (~8\% of original area), creating increasingly stringent tests of reconstruction fidelity under progressive miniaturization.

\subsection{Implementation details}

We train the network in two sequential stages. First, the local patch-wise learning deconvolution network is trained using the L1 loss function with an initial learning rate of $1 \times 10^{-1}$, decreasing by a factor of 0.1 at epochs 50, 100, and 150, over 200 total epochs. Second, the pre-trained deconvolution module is frozen and the enhancement network is trained independently with the same L1 loss at an initial learning rate of $1 \times 10^{-4}$, reduced by 0.1 at epoch 100, over 150 epochs. Both networks are implemented in PyTorch and optimized using Adam. All testings are conducted on an NVIDIA GeForce RTX 4070 Laptop GPU with 8GB memory. Patch configuration is dataset-dependent: DiffuserCam uses a $5 \times 6$ partition to accommodate its lower vertical resolution, while PhlatCam uses a $5 \times 5$ configuration for its square images, balancing reconstruction accuracy with computational efficiency.

\subsection{Experimental results}

\subsubsection{Preliminary experiment: determination of local patch number}

\begin{figure}[h]
  \centering
  \includegraphics[width=13.3cm]{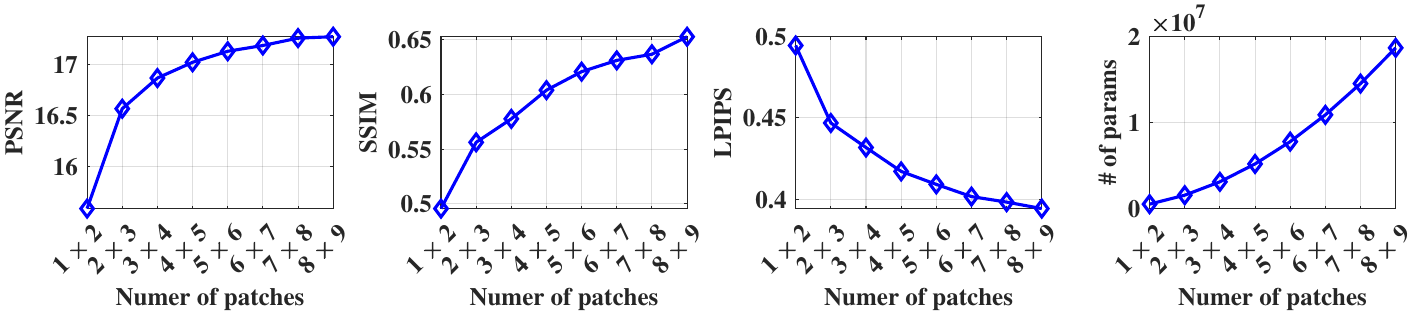}
  \caption{Relationship among reconstruction quality (PSNR, SSIM, LPIPS), parameter count, and patch number for patch-wise learning deconvolution.}
  \label{fig:deconv_curve}
\end{figure}

Before comparing our method to existing approaches, we first determine the optimal number of local patches. This preliminary study on the DiffuserCam dataset examines how patch count affects reconstruction quality. Given the dataset's non-square aspect ratio (lower vertical than horizontal resolution), we adopt an $N \times (N+1)$ partitioning strategy, allocating one additional patch horizontally. We evaluate configurations $N = 1, 2, \dots, 8$ across PSNR, SSIM, LPIPS~\cite{2018-lpips}, and parameter count. Figure~\ref{fig:deconv_curve} shows consistent quality improvements as patch number increases, since finer partitioning better approximates the spatially varying PSF within each region. However, quality gains saturate at higher patch counts while parameters scale exponentially. Balancing efficiency and accuracy, we select a $5 \times 6$ configuration for subsequent experiments, achieving satisfactory performance without prohibitive training costs.

\subsubsection{Preliminary experiment: evaluation of  local patch-wise learning deconvolution}

\begin{figure}
  \centering
  \includegraphics[width=10.5cm]{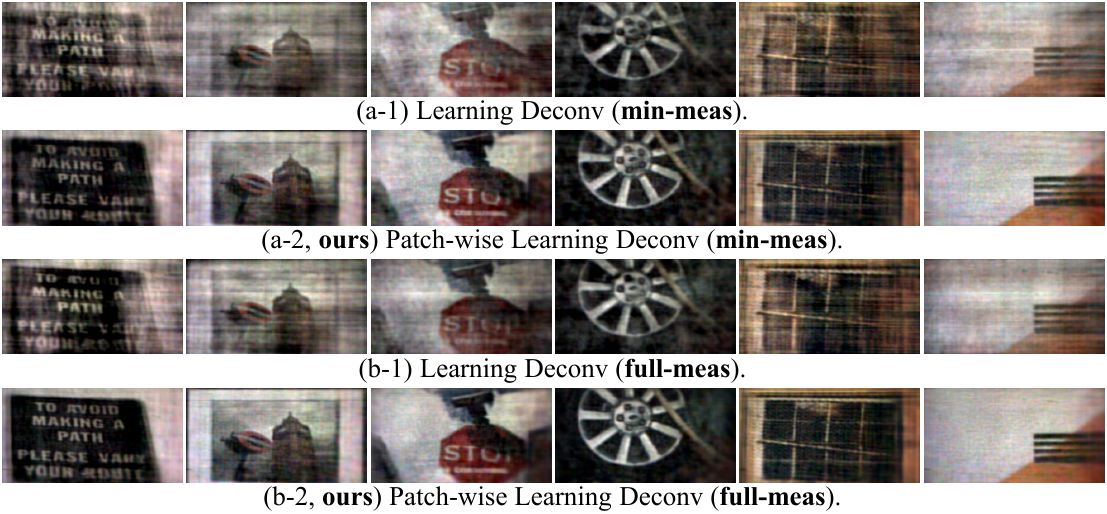}
  \caption{Global-wise v.s. patch-wise learning deconvolution on DiffuserCam dataset.}
  \label{fig:deconv_demo_diffusercam}
\end{figure}

\begin{figure}
  \centering
  \includegraphics[width=10.5cm]{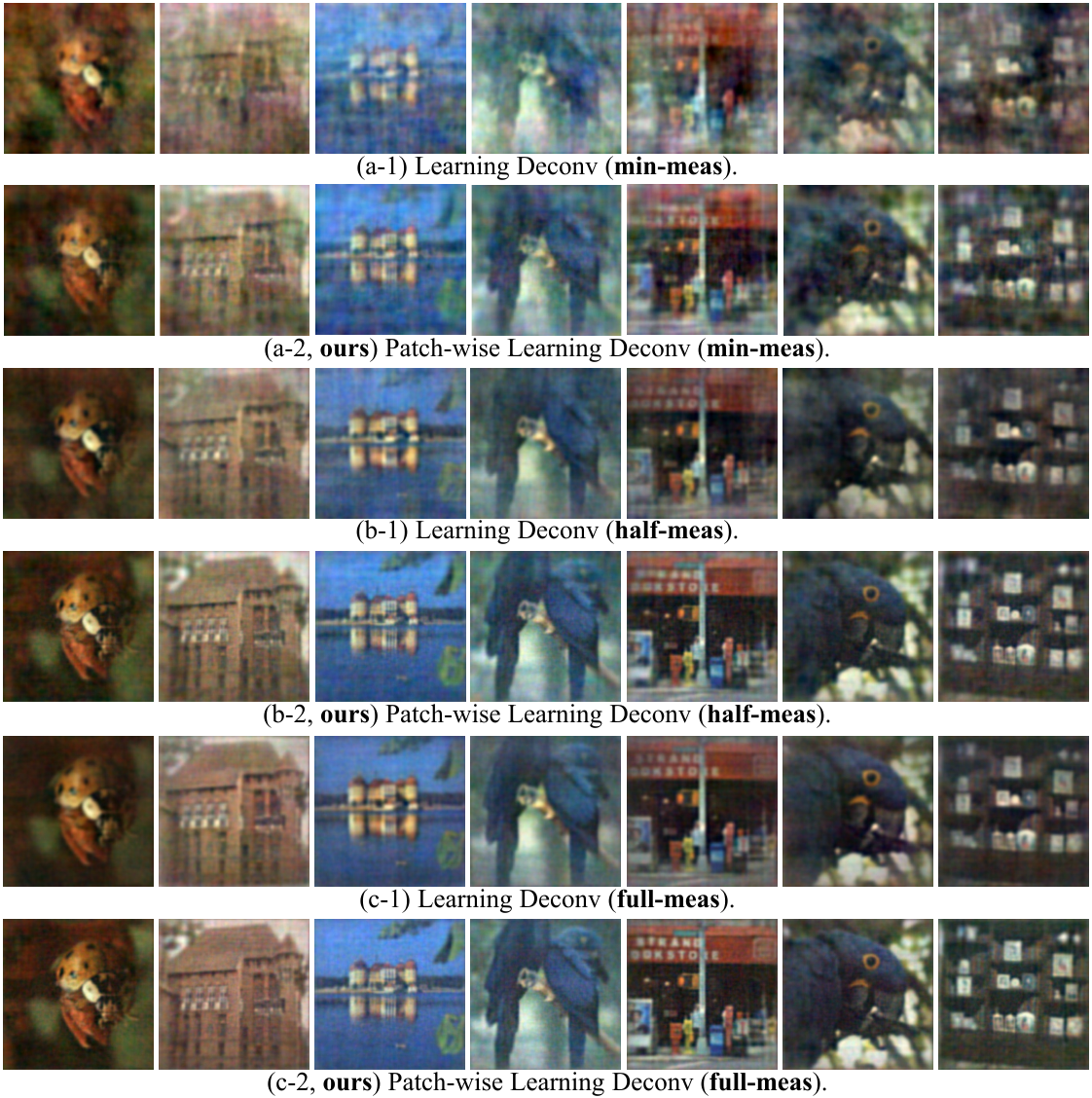}
  \caption{Global-wise v.s. patch-wise learning deconvolution on PhlatCam dataset.}
  \label{fig:deconv_demo_phlatcam}
\end{figure}

After determining the optimal patch number, we evaluate the locally stationary convolutional model and its corresponding patch-wise learning deconvolution network—the first stage of our reconstruction pipeline. We compare this approach against the conventional globally convolutional model with learning-based deconvolution~\cite{2022-flatnet} on both DiffuserCam and PhlatCam datasets under three measurement conditions: full-meas (all measurements), half-meas (primary energy region only), and min-meas (central region matching ground-truth footprint, discarding peripheral information). Quantitative results (Table \ref{tab:deconv_all}) and visual comparisons (Figures~\ref{fig:deconv_demo_diffusercam} and \ref{fig:deconv_demo_phlatcam}) demonstrate consistent superiority of the patch-wise method across all metrics (PSNR, SSIM, LPIPS), yielding sharper reconstructions with substantially reduced noise and artifacts.

Detailed analysis reveals substantial improvements across both datasets. On DiffuserCam, the patch-wise method produces clearer car wheel spokes with cleaner backgrounds, sharper text with reduced distortion, and fewer color artifacts in dense patterns. On PhlatCam, ladybird wing spots appear brighter, bookshelf and brick wall edges sharper, complex mixed scenes more faithful, and textured regions—water ripples and feathers—notably enhanced. Most remarkably, under DiffuserCam's stringent min-meas condition where measurement truncation is most severe, the patch-wise method achieves reconstruction clarity substantially exceeding that of the conventional method operating with complete measurements (Figure~\ref{fig:deconv_demo_diffusercam}, rows 2–3), demonstrating that the locally stationary convolutional model effectively compensates for information loss.

\begin{table}[h]
  \caption{Comparison between patch-wise and global-wise learning deconvolution.}
  \label{tab:deconv_all}
  \vspace{-10pt}
  \centering
  \footnotesize
  \begin{tabular}{cc|ccc|ccc}
    \hline
    \multirow{2}{*}{Dataset} & \multirow{2}{*}{Meas.} 
      & \multicolumn{3}{c|}{Global-wise Learning Deconv} 
      & \multicolumn{3}{c}{Patch-wise Learning Deconv} \\
    \cline{3-8}
    & & PSNR$\uparrow$ & SSIM$\uparrow$ & LPIPS$\downarrow$
      & PSNR$\uparrow$ & SSIM$\uparrow$ & LPIPS$\downarrow$ \\
    \hline
    \multirow{2}{*}{DiffuserCam} 
      & Full & 15.2631 & 0.4635 & 0.5332 & 17.1315 & 0.6207 & 0.4091 \\
      & Min  & 13.4382 & 0.4145 & 0.5966 & 14.8599 & 0.5088 & 0.5403 \\
    \hline
    \multirow{3}{*}{PhlatCam} 
      & Full & 15.3735 & 0.4784 & 0.6541 & 16.2941 & 0.4864 & 0.6201 \\
      & Half & 15.0071 & 0.4627 & 0.6905 & 16.2720 & 0.4863 & 0.6545 \\
      & Min  & 14.8011 & 0.4347 & 0.7438 & 15.9780 & 0.4623 & 0.7119 \\
    \hline
  \end{tabular}
\end{table}

\newpage
\subsubsection{Experiment: comparison with representative methods}

\begin{figure}[t]
\centering
\includegraphics[width=11.5cm]{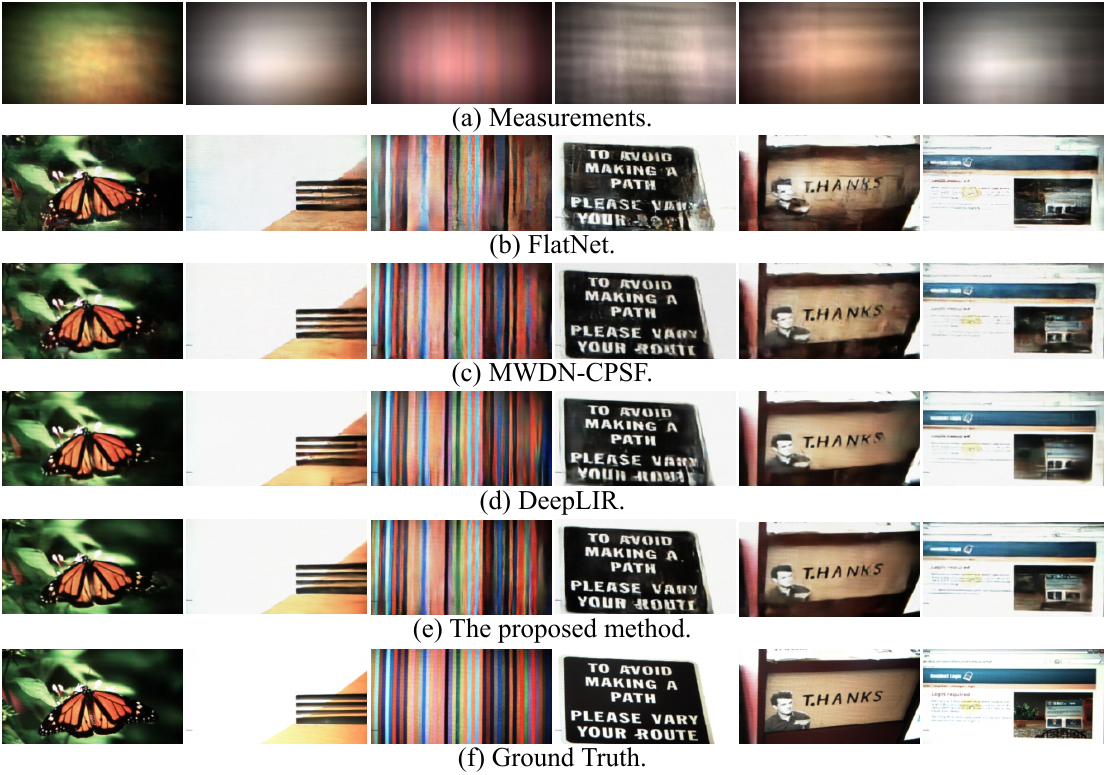}
\caption{Evaluation of different methods (DiffuserCam dataset under \textbf{full-meas} condition).}
\label{fig:result_diffusercam_full_meas}
\end{figure}

\begin{figure}[t]
\centering
\includegraphics[width=11.5cm]{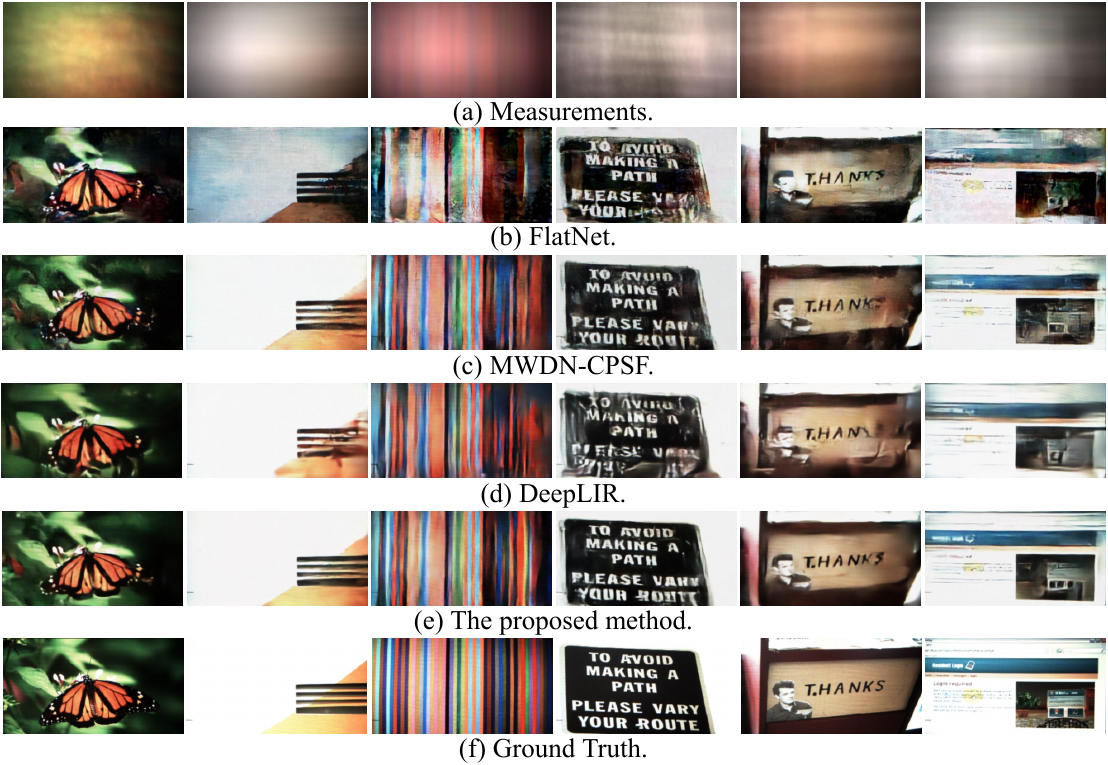}
\caption{Evaluation of different methods (DiffuserCam dataset under \textbf{min-meas} condition).}
\label{fig:result_diffusercam_min_meas}
\end{figure}

\begin{figure}[t]
  \centering
  \includegraphics[width=9.5cm]{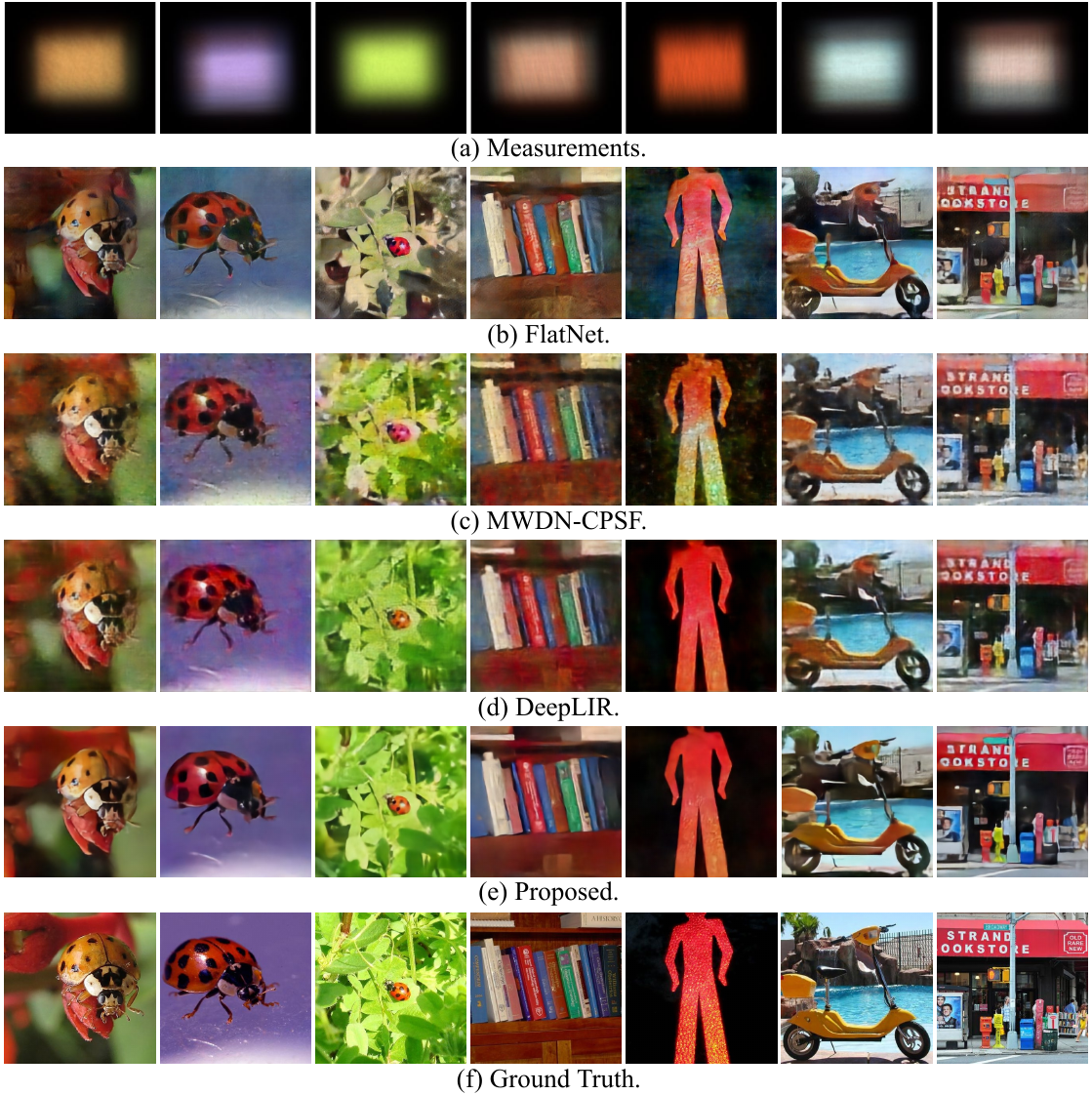}
  \caption{Evaluation of different methods (PhlatCam dataset under \textbf{full-meas} condition).}
  \label{fig:result_phlatcam_full_meas}
\end{figure}

\begin{figure}[t]
  \centering
  \includegraphics[width=9.5cm]{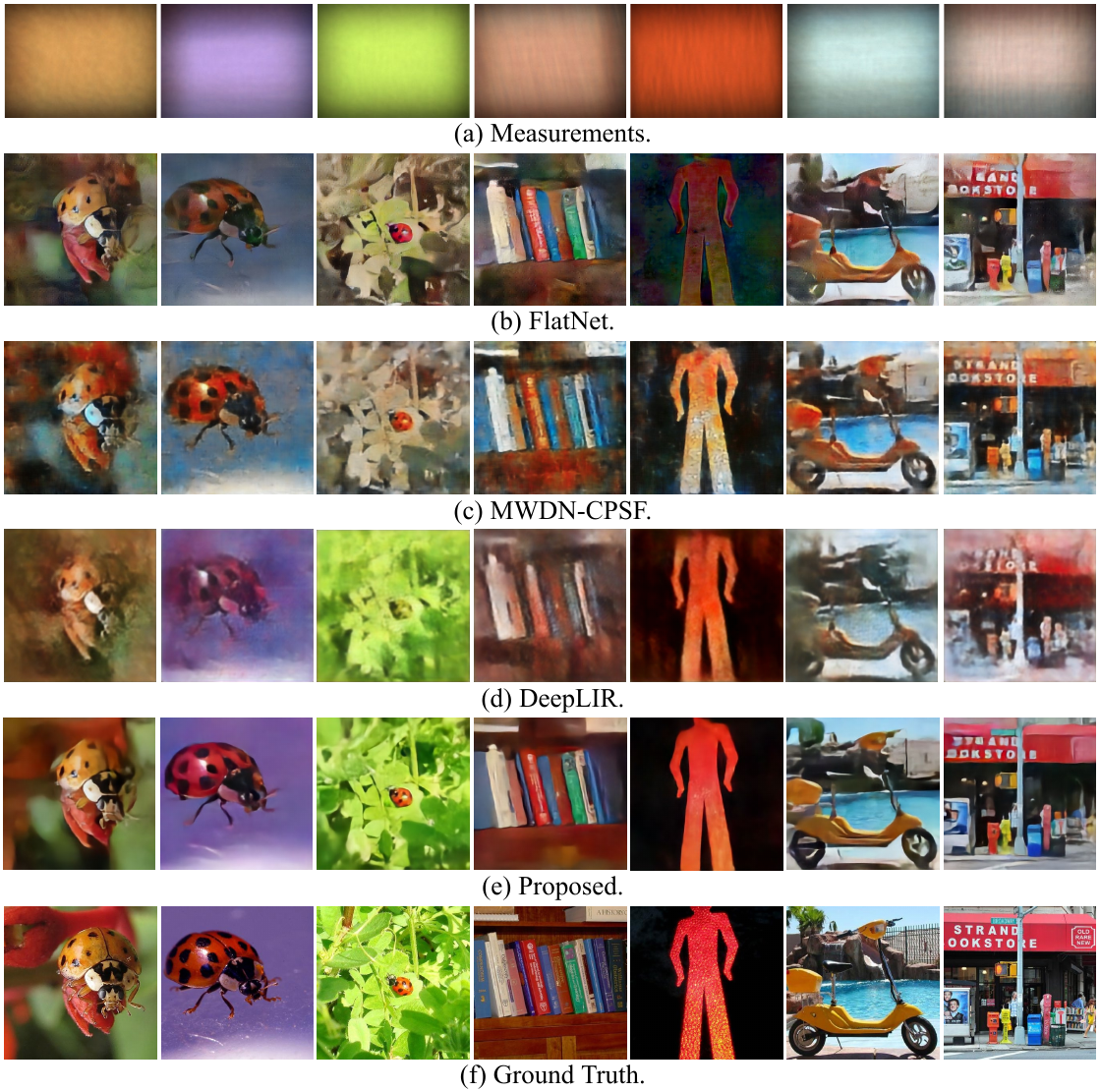}
  \caption{Evaluation of different methods (PhlatCam dataset under \textbf{half-meas} condition).}
  \label{fig:result_phlatcam_half_meas}
\end{figure}

\begin{figure}[t]
  \centering
  \includegraphics[width=9.5cm]{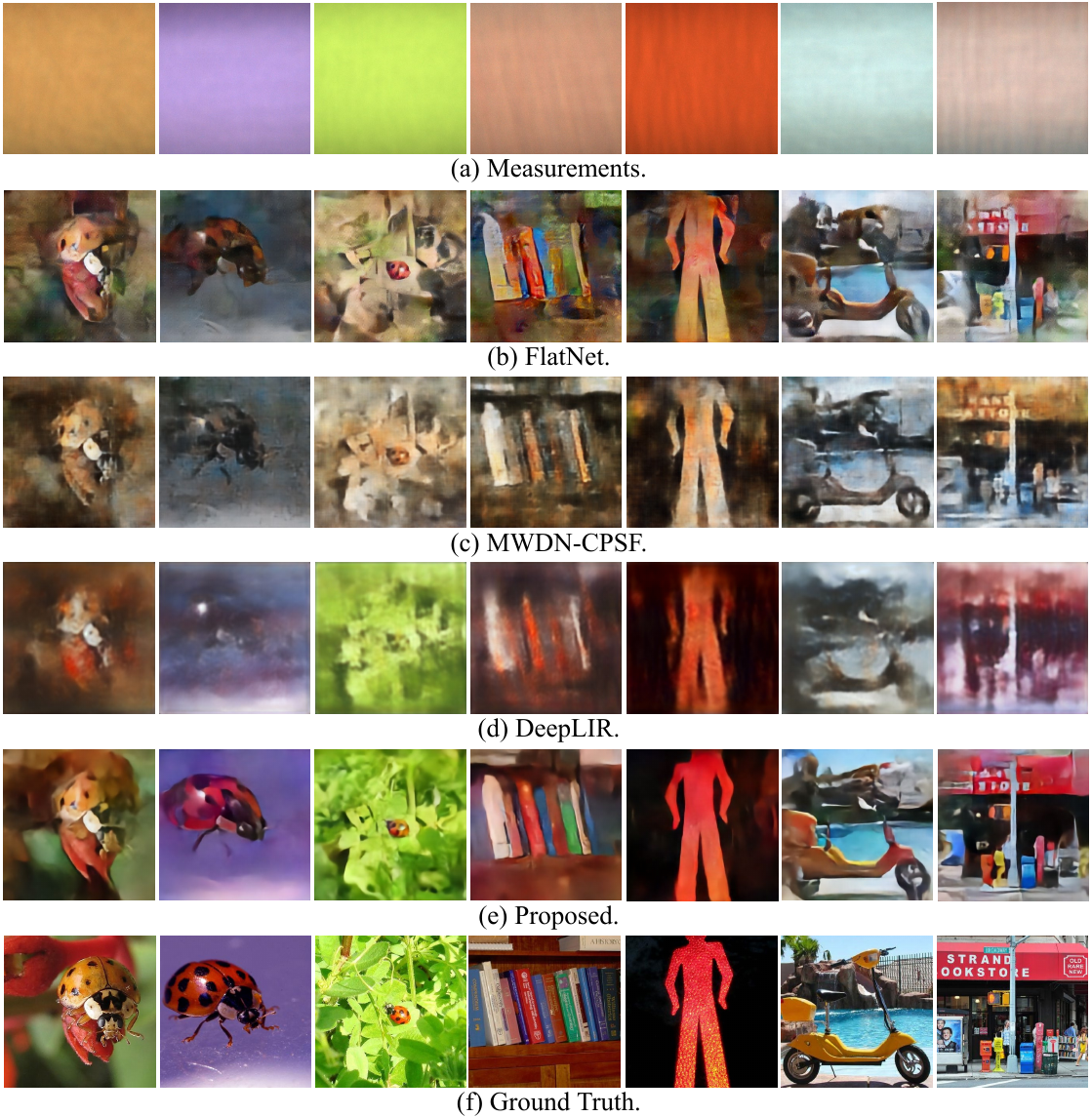}
  \caption{Evaluation of different methods (PhlatCam dataset under \textbf{min-meas} condition).}
  \label{fig:result_phlatcam_min_meas}
\end{figure}

Having validated our core hypothesis, we now benchmark the complete local-to-global hierarchical method against three representative techniques—FlatNet (two-stage), DeepLIR (deep unfolding), and MWDN (multi-scale). Quantitative metrics and visual comparisons for DiffuserCam appear in Table~\ref{tab:result:diffusercam} and Figures~\ref{fig:result_diffusercam_full_meas} (full-meas) and \ref{fig:result_diffusercam_min_meas} (min-meas); PhlatCam results are shown in Table~\ref{tab:result:phlatcam} and Figures~\ref{fig:result_phlatcam_full_meas} (full-meas), \ref{fig:result_phlatcam_half_meas} (half-meas), and \ref{fig:result_phlatcam_min_meas} (min-meas).

Quantitative evaluation demonstrates that our method consistently outperforms the three comparative approaches across almost all metrics under identical experimental conditions. Qualitative analysis reveals substantially fewer artifacts and notably higher fidelity in recovering fine-scale structures, especially under severe measurement truncation and near image boundaries.

On DiffuserCam (Figures~\ref{fig:result_diffusercam_full_meas} and \ref{fig:result_diffusercam_min_meas}), our method achieves superior fidelity: stripe patterns (columns two and three) closely match ground-truth intensity profiles, text-heavy scenes (columns four and five) exhibit enhanced clarity with reduced background artifacts, butterfly wing edges (column one) show improved definition, and the rightmost column displays sharper frame boundaries with clearer internal details. On PhlatCam (Figures~\ref{fig:result_phlatcam_full_meas}, \ref{fig:result_phlatcam_half_meas}, and \ref{fig:result_phlatcam_min_meas}), improvements include smoother textures and more accurate color reproduction (columns one, two, three, and five), more realistic plant leaves with faithful texture rendering (column three), clearer inter-book boundaries in the bookshelf scene (column four), and reduced geometric distortion with fewer artifacts in fine-detail regions (rightmost columns). These results confirm that the local-to-global hierarchical framework effectively addresses spatial PSF variation and measurement truncation across diverse imaging conditions.

As measurement size decreases from full-meas to min-meas, our method exhibits substantially greater robustness than existing approaches. On DiffuserCam, reconstruction quality remains nearly constant across all conditions, maintaining high fidelity even under severe truncation, while comparative methods show marked degradation in typographic details, stripe patterns, and edge structures. PhlatCam results demonstrate similar resilience: half-meas reconstructions are perceptually indistinguishable from full-meas, whereas reference methods suffer noticeable quality loss under equivalent constraints. Most remarkably, under extreme min-meas conditions where comparative methods fail to achieve adequate structural reconstruction, our approach continues to recover primary image content effectively. The proposed method also maintains computational efficiency with modest parameter count (the rightmost column of Table \ref{tab:result:diffusercam}) and consistently faster reconstruction times than most alternatives.

\subsection{Ablation study}

To systematically validate each component's contribution, we have conducted an ablation study on the DiffuserCam dataset under full-meas conditions, evaluating four configurations: (1) global-wise conventional learning deconvolution with standard UNet enhancement~\cite{2015-unet}; (2) global-wise deconvolution with our proposed enhancement network; (3) patch-wise learning deconvolution with standard UNet enhancement; and (4) the complete proposed method combining patch-wise deconvolution with our hierarchical enhancement network.

Results in Table~\ref{tab:ablation_study} demonstrate that the complete method achieves substantial improvements over baseline: 4.79 dB PSNR gain, 13.4\% SSIM enhancement, and 18.1\% LPIPS reduction. Patch-wise deconvolution emerges as the primary contributor, delivering 3.2 dB PSNR and 7\% SSIM improvements by effectively addressing spatial PSF variations. The hierarchical enhancement network provides additional refinement, contributing 1.6 dB PSNR and 3\% SSIM gains while reducing LPIPS from 0.14 to 0.13 through progressive detail recovery. Notably, despite utilizing only 28 million parameters—substantially fewer than the 124M–132M required by alternative configurations—the complete system achieves superior performance, demonstrating that physics-informed architectural design enables both computational efficiency and reconstruction quality.

\clearpage

\begin{table}[t]
  \caption{Evaluation of different methods on the DiffuserCam dataset.}
  \label{tab:result:diffusercam}
  \centering
  \scriptsize
  \renewcommand{\arraystretch}{1.2} 
  \begin{tabular}{c|c@{\hspace{4pt}}c@{\hspace{4pt}}|c@{\hspace{4pt}}c@{\hspace{4pt}}|c@{\hspace{4pt}}c@{\hspace{4pt}}|c@{\hspace{4pt}}c@{\hspace{4pt}}|c}
    \hline
    \multirow{2}{*}{Method} 
      & \multicolumn{2}{c@{\hspace{4pt}}|}{PSNR$\uparrow$} 
      & \multicolumn{2}{c@{\hspace{4pt}}|}{SSIM$\uparrow$} 
      & \multicolumn{2}{c@{\hspace{4pt}}|}{LPIPS$\downarrow$} 
      & \multicolumn{2}{c@{\hspace{4pt}}|}{Time (s)} 
      & \multirow{2}{*}{\# of params} \\
    \cline{2-9}
      & Full-meas & Min-meas 
      & Full-meas & Min-meas 
      & Full-meas & Min-meas 
      & Full-meas & Min-meas 
      & \\
    \hline
    FlatNet   & 22.80 & 17.39 & 0.77 & 0.52 & 0.16 & 0.29 & 0.21 & 0.17 & 124M \\
    MWDN      & 27.09 & 22.52 & 0.85 & 0.71 & 0.13 & 0.19 & 0.12 & 0.10 & 21M  \\
    DeepLIR   & 26.65 & 19.86 & 0.85 & 0.62 & 0.14 & 0.30 & 0.22 & 0.14 & 18M  \\
    Proposed  & 27.59 & 23.33 & 0.87 & 0.75 & 0.13 & 0.21 & 0.15 & 0.13 & 28M  \\
    \hline
  \end{tabular}
\end{table}

\clearpage

\clearpage
\begin{table}[t]
  \caption{Evaluation of different methods on the PhlatCam dataset.}
  \centering
  \footnotesize
  \begin{tabular}{c|ccccc}
    \hline
    & \textbf{Method} & \textbf{PSNR}$\uparrow$ & \textbf{SSIM}$\uparrow$ & \textbf{LPIPS}$\downarrow$ & \textbf{Time (s)} \\
    \hline
    \multirow{4}{*}{\shortstack{Full meas}}
    & FlatNet & 19.03 & 0.55 & 0.31 & 0.20 \\
    & MWDN    & 21.17 & 0.59 & 0.42 & 0.38 \\
    & DeepLIR & 20.82 & 0.58 & 0.46 & 0.40 \\
    & Proposed & 22.07 & 0.62 & 0.36 & 0.21 \\
    \hline
    \multirow{4}{*}{\shortstack{Half meas}}
    & FlatNet & 16.88 & 0.49 & 0.39 & 0.19 \\
    & MWDN    & 18.84 & 0.55 & 0.48 & 0.18 \\
    & DeepLIR & 17.23 & 0.50 & 0.59 & 0.20 \\
    & Proposed & 21.23 & 0.60 & 0.39 & 0.14 \\
    \hline
    \multirow{4}{*}{\shortstack{Min meas}}
    & FlatNet & 15.15 & 0.43 & 0.46 & 0.19 \\
    & MWDN    & 16.52 & 0.47 & 0.57 & 0.15 \\
    & DeepLIR & 16.02 & 0.46 & 0.68 & 0.18 \\
    & Proposed & 18.77 & 0.53 & 0.48 & 0.14 \\
    \hline
  \end{tabular}
  \label{tab:result:phlatcam}
\end{table}

\clearpage
\begin{table}[t]
  \caption{Ablation study of the proposed method on the DiffuserCam dataset.}
  \centering
  \footnotesize
  \begin{tabular}{l|cccc}
    \hline
    & \textbf{PSNR}$\uparrow$ & \textbf{SSIM}$\uparrow$ & \textbf{LPIPS}$\downarrow$ & \textbf{\# of params} \\
    \hline
    Global-wise Learning Deconv + UNet & 22.8041 & 0.7650 & 0.1558 & 124M \\
    Global-wise Learning Deconv + Proposed Enhancement & 24.9381 & 0.8319 & 0.1669 & 21M \\
    Patch-wise Learning Deconv + UNet & 26.0086 & 0.8349 & 0.1438 & 132M \\
    Proposed & 27.5916 & 0.8675 & 0.1275 & 28M \\
    \hline
  \end{tabular}
  \label{tab:ablation_study}
\end{table}

\section{Conclusion}

This work addresses fundamental limitations in lensless imaging systems that arise when pursuing large effective fields of view with compact sensors. Two critical factors constrain performance: the spatially varying point spread function and measurement truncation imposed by finite sensor dimensions. We introduce a locally stationary convolutional model that accurately characterizes these spatial dependencies and truncation effects, coupled with a local-to-global hierarchical reconstruction framework that preserves imaging fidelity even under substantial sensor miniaturization. The framework operates in two stages. First, patch-wise learning deconvolution employs adaptive local kernels to capture spatial variance and mitigate truncation artifacts within each image region. Second, hierarchical enhancement synthesizes these local reconstructions into a globally coherent result, refining structural consistency across the full field of view while maintaining fine-scale detail.

Extensive validation on the DiffuserCam and PhlatCam datasets demonstrates superior reconstruction quality across diverse measurement conditions, effectively increasing the usable field of view. Remarkably, under severe truncation scenarios, our method surpasses conventional approaches that operate with complete measurements. These findings remove critical barriers to ultra-compact lensless imaging systems, opening new possibilities in resource-constrained environments such as embedded vision, wearable devices, and distributed sensor networks.
\appendix
\section{Derivation of formulation (\ref{eq:imaging_model})}

As the effect introduced by noise and window function $\mathcal{W}$ are mainly addressed in the enhancement stage, whereas the first stage is devoted solely to reconstructing the latent image from the measurement domain, both noise and windowing function are neglected in the present derivation to simplify the analysis, yielding the model

\[
\mathbf{Y} = \sum_{b=1}^{B_x \times B_y} \mathbf{H}_b \circledast \mathbf{X}_b,
\]

whose frequency-domain representation is

\[
\mathcal{F}(\mathbf{Y}) = \sum_{b=1}^{B_x \times B_y} \mathcal{F}(\mathbf{H}_b) \odot \mathcal{F}(\mathbf{X}_b).
\]

To recover the $b$-th image patch $\mathbf{X}_b$, we design a frequency-domain filter $w_b$ that acts on the measured spectrum $\mathcal{F}(\mathbf{Y})$ to produce

\[
\begin{split}
  \nonumber
  \hat{\mathbf{X}}_b &= \mathcal{F}^{-1} \left( w_b \odot \mathcal{F}(\mathbf{Y}) \right) \\ 
  &= \mathcal{F}^{-1} \left( w_b \odot \left[ \sum_{b=1}^{B_x \times B_y} \mathcal{F}(\mathbf{H}_b) \odot \mathcal{F}(\mathbf{X}_b) \right] \right) \\
  &= \mathcal{F}^{-1} \left( \sum_{b=1}^{B_x \times B_y} w_b \odot \mathcal{F}(\mathbf{H}_b) \odot \mathcal{F}(\mathbf{X}_b) \right), \text{for} \enspace b = 1, \ldots, B_x \times B_y.
\end{split}
\]

Ideally, the filter $w_b$ should satisfy two conditions: faithful reconstruction of the target patch $b$ ($w_b \odot \mathcal{F}(\mathbf{H}_b) \approx 1$) and suppression of all other patches ($w_b \odot \mathcal{F}(\mathbf{H}_j) \approx 0$ for $j \neq b$). Such a filter would isolate patch $b$ from the measurement, enabling accurate reconstruction. However, in lensless imaging systems, PSFs at different spatial locations are highly similar, causing substantial spectral overlap and severe aliasing. Consequently, no filter can strictly satisfy both conditions simultaneously, rendering exact frequency-domain decoupling unattainable.

Fortunately, strict adherence to the second condition is unnecessary. Since the final image is assembled by stitching individually reconstructed patches, each estimate $\hat{\mathbf{X}}_b$ contributes only within its designated spatial region. Errors from neighboring patches are naturally excluded during stitching and do not propagate into the final result. Introducing a binary spatial window $\mathbf{1}_b$ that selects the region corresponding to patch $b$, the complete reconstruction becomes:

\[
\hat{\mathbf{X}} = \sum_{b=1}^{B_x \times B_y} \mathbf{1}_b \odot \hat{\mathbf{X}}_b.
\]

This formulation clarifies why approximate solutions remain effective: although inter-patch interference persists in the frequency domain, spatial localization ensures that only region-specific information is retained. Given the limited spatial variation of PSFs in lensless systems, a filter derived from the global PSF already provides a reasonable approximation for the mapping $\mathbf{Y} \to \hat{\mathbf{X}}_b$. When learned in a data-driven manner, such a filter minimizes reconstruction error in a global sense. The patch-wise strategy refines this global solution locally, tailoring each filter $w_b$ to its specific patch and thereby achieving finer-grained optimization and improved reconstruction fidelity.

\vspace{0.5em}
\textbf{Funding.} National Natural Science Foundation of China under Grants 62471113, 62305049, and 62371104; Sichuan Science and Technology Support Program under Grant 2024NSFSC1439.

\vspace{0.5em}
\textbf{Disclosures.} The authors declare that there are no conflicts of interest.
\vspace{0.5em}

\textbf{Data availability.} All data needed to evaluate the conclusions are present in the paper's linked public datasets.
\vspace{0.5em}


\bibliography{sample}






\end{document}